\documentclass{aa}
\usepackage{graphicx}

\def\ltsim{\raise 2pt \hbox {$<$} \kern-1.1em \lower 4pt \hbox {$\sim$}}
\def\ltapprox{\raise 2pt \hbox {$<$} \kern-1.1em \lower 5pt \hbox {$\approx
$}}
\def\gtsim{\raise 2pt \hbox {$>$} \kern-1.1em \lower 4pt \hbox {$\sim$}}
\def\gtapprox{\raise 2pt \hbox {$>$} \kern-1.1em \lower 5pt \hbox {$\approx
$}}

\begin{document}

\title{Tailed radio galaxies as tracers of galaxy clusters. 
Serendipitous discoveries with the GMRT}
\author{S.~Giacintucci\inst{1,}\inst{2} \and
T.~Venturi\inst{1}}
\institute
{
INAF -- Istituto di Radioastronomia, via Gobetti 101, I-40129, Bologna, Italy 
\and
Harvard--Smithsonian Centre for Astrophysics, 
60 Garden Street, Cambridge, MA 02138, USA
}

\date{Received 00 - 00 - 0000; accepted 00 - 00 - 0000}

\titlerunning{Tailed radio galaxies as tracers of galaxy clusters}
\authorrunning{Giacintucci \& Venturi}

\abstract
{}
{We report on the discovery of four new radio galaxies with tailed morphology.
Tailed radio galaxies are generally found in rich environments, therefore 
their presence can be used as tracer of a cluster.} 
{The radio galaxies were found in the fields of Giant Metrewave Radio 
Telescope (GMRT) observations carried out at 610 MHz and 327 MHz devoted to
other studies. We inspected the literature and archives in the optical 
and X--ray bands to search for galaxy clusters or groups hosting them.}
{All the tailed radio galaxies serendipitously found in the GMRT fields 
are located in rich environments. Two of them belong to the candidate cluster
NCS\,J090232+204358, located at z$_{\rm phot}$=0.0746; one belongs to the
cluster MaxBCGJ\,223.97317+22.15620 at z$_{\rm phot}$=0.2619; finally we 
suggest that the fourth one is probing a galaxy cluster at z=0.1177, located 
behind A\,262, and so far undetected in any band.
Our results strenghten the relevance of high sensitivity and high resolution
radio data in the detection of galaxy clusters at intermediate
redshift.}
{}
\maketitle
\section{Distorted radio sources in clusters of galaxies}\label{sec:intro}

Radio galaxies located in dense environments, such as galaxy clusters, 
often show complex and prominently distorted radio structures. A common 
morphology is represented by tailed radio galaxies, i.e., double sources 
with FR--I or FR-I/FR--II morphology (Fanaroff \& Riley 1974), whose jets 
and lobes are bent in U or C shapes. In wide--angle--tail radio galaxies 
(WAT, or C--shaped) the angle formed by the jets and lobes is usually very 
large (e.g. O'Donoghue et al. 1993; Feretti \& Venturi 2002 for a review). 
The prototype of this class is 3C\,465 in A\,2634 (Eilek et al. 1984). 
Sources whose jets and lobes form a small angle are referred to as 
narrow--angle--tail radio galaxies (NAT, or U--shaped; see Bliton et 
al. 1998, Feretti \& Venturi 2002 and references therin). The prototype 
of this class is NGC\,1265 in the Perseus cluster (Wellington et al. 1973; 
O'Dea \& Owen 1986). 

There is general consensus that the origin of the distorted radio 
morphologies lies in the interaction of the radio galaxy with 
respect to the dense surrounding intracluster medium (ICM), even though 
the details of such interaction are not fully disentangled.
NAT sources are usually associated with galaxies moving at high velocity 
in the gravitational potential of the cluster (Miley 1980). 
The ram pressure exerted by the external medium is expected to be large 
enough to curve the jets and sweep the radio emitting plasma behind the 
rapidly moving galaxy. In agreement with this idea, the optical counterparts 
of narrow--angle--tails are not the bright dominant D or cD galaxies, usually 
nearly at rest in the cluster potential well, but rather galaxies with 
fainter optical magnitude and lower radio luminosity (Rudnick \& Owen 
1976; Valentijn 1979). 
However, the host galaxies of NATs have, on average, velocities 
similar to those of typical cluster members, rather than moving with the 
high peculiar velocity expected in the ram pressure scenario. Therefore, 
it has been suggested that the jet bending might be a by--product, at least 
partially, of bulk motions in the ICM induced by cluster--subcluster mergers 
(Bliton et al. 1998).

The physical mechanism responsible for the distorsion of the jets in WATs is 
not completely understood either. Both ram pressure (Owen \& Rudnick 
1976; Begelman et al. 1979) and buoyancy forces, active if the 
jet density is lower than the external gas density (Burns \& Balonek 1982), 
may play a role, the latter being dominant at larger distances from the cluster
centre (Sakelliou et al. 1996). 
However, the very low peculiar velocities of their optical 
counterparts (Bird 1994; Pinkney et al. 2000) -- the cluster dominant D or cD 
ellipticals -- is not consistent with the observed curvatures. The presence of 
large scale bulk flows in the ICM, induced by cluster or group mergers, is an 
appealing alternative for the jet bending (Pinkney et al. 1994; 
Roettiger et al. 1996; Gomez et al. 1997). Finally, a connection between the 
radio bending and gas sloshing in cluster cores has recently been suggested 
(e.g. A\,2029; Clarke et al. 2004; Ascasibar \& Markevitch 2006).

%
%
\begin{figure*}
\centering
\includegraphics[angle=0,width=17cm]{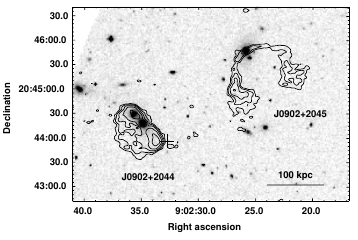}
\caption{GMRT 610 MHz contours of the two tailed radio galaxies 
discovered at $\sim$27 $^{\prime}$ from the galaxy cluster Z\,2089. 
The image is corrected for the primary beam. The radio contours are 
overlaid on the red optical image from the SDSS. The resolution of the 
radio image is 6.5$^{\prime \prime} \times 4.5^{\prime \prime}$, p.a. 
80$^{\circ}$. The lowest contour is 0.5 mJy b$^{-1}$, and each contour 
increases by a factor of two. The cross indicates the centre of the 
candidate galaxy cluster NSC\,J090232+204358. The linear scale is 
1$^{\prime\prime}$=1.56 kpc (see Sect.~\ref{sec:opt1})}
\label{fig:wats_1}
\end{figure*}
%
%

The association between tailed radio galaxies and clusters can be used to 
trace the high density structures in the Universe. Indeed many previously 
unobserved clusters have been identified thanks to the detection of tailed 
sources (e.g. Fomalont \& Bridle 1978; Burns \& Owen 1979; and more recently 
Blanton et al. 2000, 2001 and 2003; Smol\u{c}i\'{c} et al. 2007; Kantharia 
et al. 2009). In this paper we report on the discovery with the 
Giant Metrewave Radio Telescope (GMRT) of a number of distorted radio galaxies 
and on the successful search for the galaxy clusters in which they reside.

The paper is organised as follows: in Section \ref{sec:images} we present 
the images of the new tailed radio sources; in Sections \ref{sec:opt1} and
\ref{sec:opt2} we describe the optical identification procedure and the 
association with candidate clusters with photometric redshift; a brief 
summary and conclusions are reported in Section \ref{sec:summary}.
\\
We adopt the $\Lambda$CDM cosmology, with H$_0$=70 km 
s$^{-1}$ Mpc$^{-1}$, $\Omega_m=0.3$ and $\Omega_{\Lambda}=0.7$. The 
spectral index $\alpha$ is defined according to S$\propto \nu^{-\alpha}$. 
 
\section{Discovery of new radio galaxies}\label{sec:images}

In Venturi et al. (2007 and 2008) we presented a large radio survey 
of a sample of galaxy clusters carried out with the GMRT at 610 
MHz (the {\it GMRT Radio--Halo Survey}). During the analysis of 
those observations, we discovered three tailed radio galaxies in 
the wide--field images of two clusters of the sample. In particular, 
two of them, GMRT--J\,0902+2044 and GMRT--J\,0902+2045,  
are located at $\sim27^{\prime}$ from the phase centre of the observation 
of Z\,2089; the third tailed radio source, GMRT--J\,1455+2209, 
is located at $\sim 22^{\prime}$ from the pointing of the cluster Z\,7160. 

Similarly, we found a diffuse and amorphous radio source, GMRT--J0154+3627, 
while imaging the 327 MHz emission of the cluster A\,262 (archival GMRT 
observations), as part of a large observational project devoted 
to the study of the radio source feedback in groups and poor clusters of 
galaxies (Giacintucci et al. in prep.). The source is located at 
$\sim$32$^{\prime}$ to the North--East of the antenna pointing of A\,262. 

Since the half power width of the primary beam of a GMRT antenna 
is $\sim 50^{\prime}$ at 610 MHz and $\sim$ 90$^{\prime}$ at 327 MHz, 
all the new radio sources lie well within the primary beam of the Z\,2089, 
Z\,7160 and A\,262 fields.

\subsection{GMRT radio observations}

The details on the GMRT observations are summarised in Table 
\ref{tab:obs}, where we provide the radio source name; pointing 
coordinates of the observation; distance from the pointing; 
frequency, bandwidth and total lenght of the observations; half 
power beamwidth (HPBW) of the full array; rms level (1$\sigma$) 
measured in the region of the sources, prior to the primary beam 
correction. All the GMRT observations in Tab.~\ref{tab:obs} were 
obtained using both the upper and lower side bands of 16 MHz each. 
The data were acquired in spectral line mode, with 128 channels/band 
and a spectral resolution of 125 kHz/channel. The data reduction was 
carried out using the standard procedure (calibration, Fourier inversion, 
clean and restore) with the NRAO Astronomical Image Processing System 
(AIPS) package. In order to reduce the size of the datasets and to 
minimize bandwidth smearing effects within the primary beam, after bandpass 
calibration the central channels of each observation were averaged to 6 
channels of $\sim$ 2 MHz each. After further careful editing in the 
averaged datasets, a number of phase-only self--calibration cycles and 
wide--field imaging were carried out for each band. The final images were 
produced by combining the USB and LSB self--calibrated datasets. Due to 
residual RFI in the LSB dataset of J\,0154+3627 (A\,262), the data combination 
led to images whose quality is worse than those obtained from the USB alone. 
For this reason only the USB dataset was used for the analysis presented
in this paper. Residual amplitude calibration errors are $\le 5$\%. 
We refer to Venturi et al. (2007 and 2008) and Giacintucci 
et al. (2008) for a complete description of the data reduction.

%
%
\begin{figure}
\centering
\includegraphics[angle=0,width=\hsize]{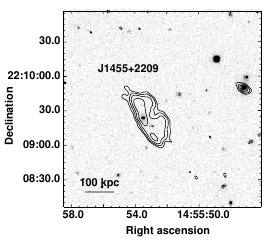}
\caption{GMRT 610 MHz contours of the WAT discovered at $\sim$22
$^{\prime}$ from the galaxy cluster Z\,7160. The image is corrected
for the primary beam. The radio contours are overlaid on the red optical 
image from the SDSS. The resolution of the radio image is 6.1$^{\prime \prime} 
\times 4.1^{\prime \prime}$, p.a. 39$^{\circ}$. The lowest contour is 0.4 mJy 
b$^{-1}$, and each contour increases by a factor of two. The centre of 
the galaxy cluster MaxBCG\,J223.97317+22.15620 is coincident with the 
elliptical galaxy associated with the WAT. The linear scale is 
$1^{\prime\prime}$=3.99 kpc (see Sect.~\ref{sec:opt1}).}
\label{fig:wats_2}
\end{figure}
%
%

\subsection{The 610 MHz images of the tailed radio galaxies}\label{sec:tailed}

In Figure \ref{fig:wats_1} we show the 610 MHz contours of 
the two tailed radio sources GMRT--J\,0902+2044 and GMRT--J\,0902+2045
(hereinafter  J\,0902+2044 and J\,0902+2045 respectively), discovered 
in the field of Z\,2089 (Tab.~\ref{tab:obs}). 
The two radio galaxies, both extending well beyond the optical size of
the associated counterpart, are separated by $\sim 2.6^{\prime}$ in the plane
of the sky. The contours are 
superposed to the red optical image from the Sloan Digital Sky Survey (SDSS;
Data Release 7\footnote{www.sdss.org$/$dr7$/$}). 
The radio image  has been corrected for the primary beam attenuation of 
the GMRT at 610 MHz. In the region shown in Fig.~\ref{fig:wats_1} the 
sensitivity is 60 $\mu$Jy b$^{-1}$, similar to the 
value achieved in the phase centre of the observations (45 $\mu$Jy b$^{-1}$; 
Venturi et al. 2008). After the primary  beam correction, the noise in the 
region of the radio galaxies increases to 150 $\mu$Jy b$^{-1}$.

J\,0902+2044 has a symmetric WAT morphology, with a central compact 
component and two bright jets marginally deflected to form an angle 
of $\sim 170^{\circ}$, at least out to $\sim$0.5$^{\prime}$ 
from the central radio peak. Then the jets abruptly change direction 
by $\sim 90^{\circ}$, and merge into a single low surface brightness 
tail extending East of the source. A compression of the radio isophotes 
is visible in the North--Western edge of the jets, suggesting that the 
pressure necessary to bend the radio emission is coming from that direction. 
\\
The radio galaxy J\,0902+2045 is characterized by twin--jets, smoothly 
curved in a C--shape. The jets appear rather symmetric in morphology and 
extent, and can be traced out to $\sim$1.5$^{\prime}$ from the central radio 
peak. The angle between the tails is $\sim$ 120$^{\circ}$ 
out to approximately 30$^{\prime \prime}$ from the centre. Then the jets 
undergo two more changes of direction, each with an angle of 
$\sim$120$^{\circ}$; after the last bend, they lose collimation and gently 
curve toward the outer regions. From a morphological point of view, the very 
smooth and complete bending of J\,0902+2045 resembles that of NAT radio 
galaxies, and indeed the source is strongly reminiscent of the prototypical 
narrow--angle--tail source NGC\,1265 in the Perseus cluster (e.g. 
O'Dea \& Owen, 1986; see their Figs.~4 and 5); at the same time, radio sources 
similar to J\,0902+2045 have also been classified in the literature as 
WAT (Smol\u{c}i\'{c} et al. 2007).

The integrated flux density at 610 MHz is S$_{\rm 610 \, MHz}=258.2\pm 12.9$ 
mJy for J\,0902+2044 and S$_{\rm 610 \, MHz}=145.6\pm7.3$ mJy for J\,0902+2045. 
Using the images from the NRAO VLA Sky Survey (NVSS; Condon et al. 1998), we 
measure a flux at 1.4 GHz of 128.9$\pm5.1$ mJy and 78.6$\pm3.1$ mJy, 
respectively. 
By comparison of these values with the 610 MHz fluxes, we obtained a spectral 
index $\alpha=0.83\pm0.11$ for the J0902+2044 and $\alpha=0.74^{+0.10}_{-0.11}$ 
for J0902+2045 (see Tab.~\ref{tab:wats_1}).
\\
\\
Fig. ~\ref{fig:wats_2} shows the 610 MHz image of the WAT GMRT--J\,1455+2209
(hereinafter J\,1455+2209), discovered at $\sim$22$^{\prime}$ from Z\,7160 
(Tab.~\ref{tab:obs}), overlaid on the SDSS red optical frame. 
The sensitivity in this region is 60 $\mu$Jy b$^{-1}$, and the noise level 
after the primary beam correction is 130 $\mu$Jy b$^{-1}$. The source has a 
morphology similar to J\,0902+2044 (Fig.~\ref{fig:wats_1}), however no 
clear compact component is visible at the source centre, and some level of 
asymmetry is observed in the jet brightness. The flux density of J\,1455+2209 
at 610 MHz and 1.4 GHz are S$_{\rm 610 \, MHz}=91.3\pm4.6$ mJy and 
S$_{\rm 1.4 \, GHz}= 49.1\pm2.0$ mJy (from the NVSS), implying 
$\alpha=0.75^{+0.10}_{-0.12}$ (see Tab.~\ref{tab:wats_1}). 

The spectral index $\alpha$ calculated for the three tailed sources 
is in good agreement with the typical values of extended active 
radio galaxies. The radio information is summarised in Tab.~\ref{tab:wats_1}. 
For the optical properties provided in the Table, we refer to 
$\S$~\ref{sec:opt1} .


\begin{table*}[h!]
\caption[]{Details of the GMRT observations.}
\begin{center}
\begin{tabular}{cccccccc}
\hline\noalign{\smallskip}
Source name & Pointing RA \& DEC & d & $\nu$ &  $\Delta \nu$ & Obs. time  & HPBW, PA  &   rms      \\ 
 (GMRT --)   &  (h,m,s \hspace{0.2cm} \& \hspace{0.15cm} $^{\circ}$, $^{\prime}$, $^{\prime \prime}$)  &  ($^{\prime}$) &  (MHz) & (MHz) & (min)  &
	    ($^{\prime \prime} \times^{\prime \prime}$, $^{\circ}$) & (mJy b$^{-1}$) \\
\hline\noalign{\smallskip}
J0902+2044, J0902+2045 & 09 00 45.9 \hspace{0.35cm} +20 55 13 & 27 &  610 &  32 &   80 & 6.5$\times$4.5, 80 & 60 \\
&&&&&&\\
J1455+2209 & 14 57 15.2 \hspace{0.35cm} +22 20 30 & 22 &   610 & 32 & 100 & 6.1$\times$4.1, 39 & 60  \\
&&&&&&\\
J0154+3627 & 01 52 50.0 \hspace{0.35cm} +36 09 00 & 32 & 327 & 16$^*$ & 270 & 12.2$\times$8.7, 87 & 440 \\
\hline\noalign{\smallskip}
\end{tabular}
\end{center}
\label{tab:obs}
$*$ The observations were carried out using a total bandwidth of 32 MHz (USB+LSB), but only the USB dataset
was used for the analysis.
\end{table*}

\begin{table*}[h!]
\caption[]{Radio and optical data of the tailed radio galaxies.}
\begin{center}
\begin{tabular}{lcccccc}
\noalign{\smallskip}
\hline\noalign{\smallskip}
Radio name & RA$_{\rm radio}$  & DEC$_{\rm radio}$ &  S$_{\rm 610\,MHz}$ & S$_{\rm 1.4\, GHz}$ 
&  $\alpha_{0.6\,\rm GHz}^{1.4\,\rm GHz}$ &  Radio \\ 
 (GMRT--)      & (h,m,s)  & ($^{\circ}$,$^{\prime}$, $^{\prime \prime}$) &  (mJy)  & (mJy)
&                                     & morphology \\
\noalign{\medskip}
Optical name & RA$_{\rm opt}$ & DEC$_{\rm opt}$ & z$_{\rm spec}$ & M$_{\rm R}$ & 
logP$_{1.4 GHz}$ & LLS  \\
 (SDSS--)     & (h,m,s)  & ($^{\circ}$,$^{\prime}$, $^{\prime \prime}$) & & & 
(W Hz$^{-1}$) & (kpc) \\
\hline\noalign{\smallskip}			      
J\,0902+2044 &  09 02 34.93 & +20 44 17.9 & 258.2$\pm$12.9 & 128.9$\pm$5.1 
& 0.83$\pm$0.11 & WAT \\ 
J\,090234.90+204417.9 &  09 02 34.90 & +20 44 18.0 & 0.0830$\pm$0.0002 & $-$22.42 &  
24.34 & 100 \\
&&&&&& \\
J\,0902+2045 & 09 02 25.89 & +20 45 46.6 & 145.6$\pm$7.3 &  78.6$\pm$3.1  
&  0.74$^{+0.10}_{-0.11}$ & NAT \\
J\,090225.86+204546.5 & 09 02 25.87 & +20 45 46.5 & 0.0820$\pm$0.0002 &   $-$22.21 & 
24.11 & 300    \\
&&&&&& \\
J\,1455+2209 &  14 55 53.51& +22 09 23.4 &  91.3$\pm$4.6 & 49.1$\pm$2.0 
& 0.75$^{+0.10}_{-0.12}$ & WAT \\
J\,145553.56+220922.3 &  14 55 53.56& +22 09 22.3 &  0.2566$\pm$ 0.0002 &  $-$22.61  & 
24.97 & 230  \\
\noalign{\smallskip}
\hline\noalign{\smallskip}
\end{tabular}
\end{center}
\label{tab:wats_1}
\end{table*}

\begin{table*}[h!]
\caption[]{Radio and optical data of the candidate NAT J\,0154+3627.}
\begin{center}
\begin{tabular}{lccccccc}
\noalign{\smallskip}
\hline\noalign{\smallskip}
Radio name & RA$_{\rm radio}$  & DEC$_{\rm radio}$ &  S$_{\rm 327\,MHz}$ & 
S$_{\rm 1.4\, GHz}$  &S$_{\rm 4.9\, GHz}$ &  $\alpha_{0.3\,\rm GHz}^{1.4\,\rm GHz}$  &  
logP$_{1.4 GHz}$  \\ 
(GMRT--) & (h,m,s)  & ($^{\circ}$,$^{\prime}$, $^{\prime \prime}$) &  (mJy) 
& (mJy) & (mJy) & (W Hz$^{-1}$) \\
\noalign{\medskip}
Optical name & RA$_{\rm opt}$ & DEC$_{\rm opt}$ & z$_{\rm spec}$ & r & M$_{\rm R}$  
&  $\alpha_{0.3\,\rm GHz}^{4.9\,\rm GHz}$ & LLS  \\
(NFP--)           & (h,m,s)  & ($^{\circ}$,$^{\prime}$, $^{\prime \prime}$) & &   &  
& & kpc \\
\hline\noalign{\smallskip}			      
J\,0154+3627 &  01 54 51.53  & +36 27 46.2 & 620.5$\pm$31.0 & 234.4$\pm$9.4 
& 132.8$\pm$5.3 & 0.66$^{+0.7}_{-0.6}$ & 24.93 \\ 
J\,015451.5+362747 & 09 02 34.92 &  +20 44 18.0 & 0.1177$\pm$0.0002 & 15.96 & 
$-$22.73  & 0.57$^{+0.3}_{-0.4}$ & 340      \\
\noalign{\smallskip}
\hline\noalign{\smallskip}
\end{tabular}
\end{center}
\label{tab:wats_2}
\end{table*}
%
%
%
\begin{figure*}
\centering
\includegraphics[angle=0,width=\hsize]{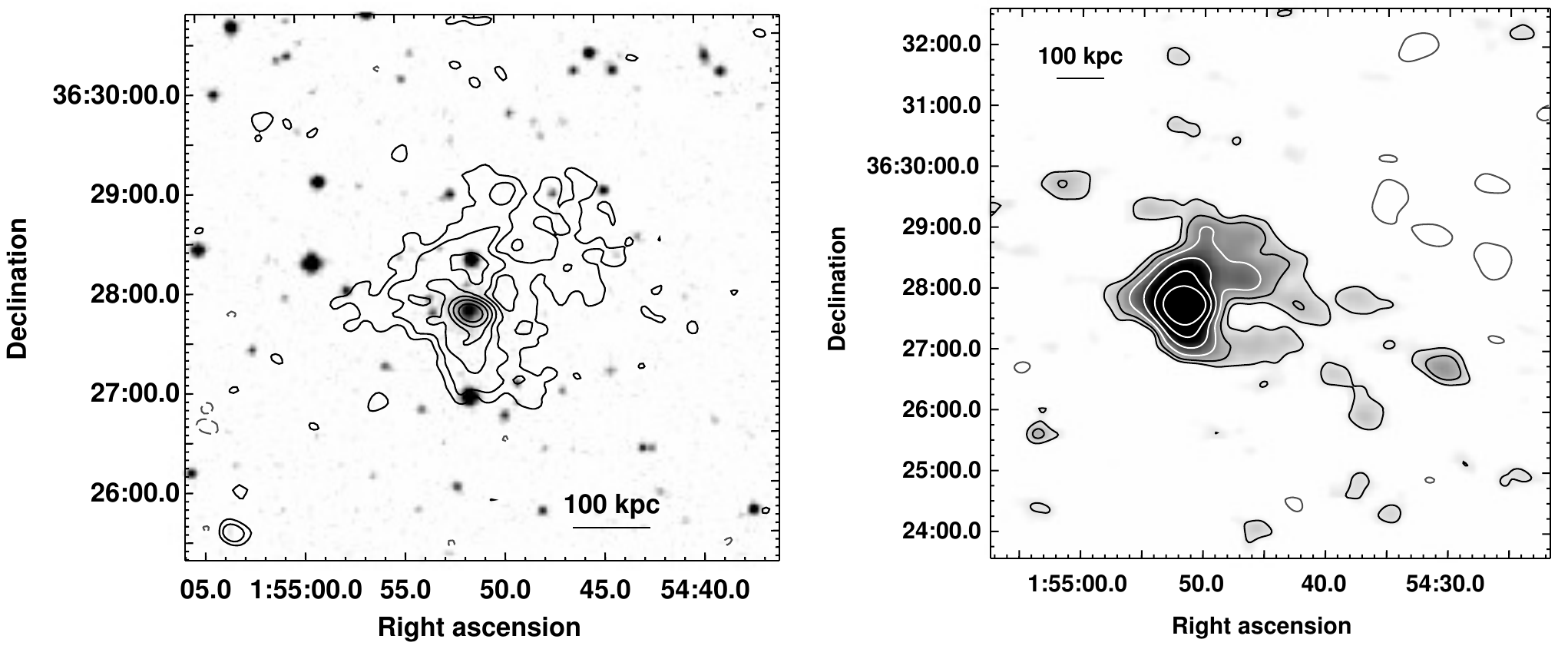}
\caption{{\it Left:} GMRT 327 MHz full resolution contours of the diffuse radio 
source J\,0154+3627, discovered at $\sim$32 $^{\prime}$ from A\,262. The image
is corrected for the primary beam attenuation. The 
radio contours are overlaid on the red optical image from the POSS--2. 
The resolution of the radio image is 12.2$^{\prime \prime} \times 8.7^{\prime \prime}$, 
p.a. 87$^{\circ}$. The lowest contour is 1.6 mJy b$^{-1}$, and each 
contour increases by a factor of two. {\it Right}: GMRT 327 MHz low
resolution image (contours and grey scale) of J\,0154+3627. The
resolution is 28.8$^{\prime \prime}\times$21.3$^{\prime \prime}$, p.a.
$-88$. Contours are spaced by a factor 2, starting from $\pm3\sigma$ (
1$\sigma=600$ mJy b$^{-1}$. For this source the scale is 
2.127 kpc/$^{\prime \prime}$ (see Sect.~4).}
\label{fig:diffuse}
\end{figure*}
%
%

\subsection{The diffuse radio source J\,0154+3627}

Fig.~\ref{fig:diffuse} shows the 327 MHz images of the diffuse radio 
source GMRT--J\,0154+3627 (hereinafter J\,0154+3627), found 
in the periphery of A\,262. In the literature this source is 
known as B2\,0151+36, being part of the Second Bologna Catalog 
(B2) of radio sources (Colla et al. 1973), obtained using the 
Bologna Northern Cross telescope at 408 MHz. However, given 
the low resolution of that instrument (3$^{\prime}$ in RA, 
and 10$^{\prime}$ in DEC), no morphological classification was 
provided at that time. The 327 MHz full resolution image, shown 
in the left panel of Fig.~\ref{fig:diffuse} overlaid on the POSS--2 red
optical plate, clearly reveals the presence 
of a bright and compact component at the source centre, two short
symmetric jets (extended $\sim 1.5^{\prime}$), and a region of low 
brightness radio emission, with a rather amorphous morphology and a size 
of $\sim2.5^{\prime}\times3^{\prime}$.
To better highlight the distribution of the low surface brightness 
emission, we produced a low resolution image at 327 MHz, shown 
in the right panel of Fig.~\ref{fig:diffuse}. At this lower resolution 
(28.8$^{\prime\prime}\times21.3^{\prime\prime}$) the source reveals 
two faint radio tails, bent toward the same south--western direction.
These tails can be well traced out to a distance of $\sim$3$^{\prime}$ 
from the compact component. However, positive residuals of emission suggest 
that they might be more even extended than what is detected at the sensitivity 
level of the image.

\begin{figure}
\centering
\includegraphics[angle=0,width=8.5cm]{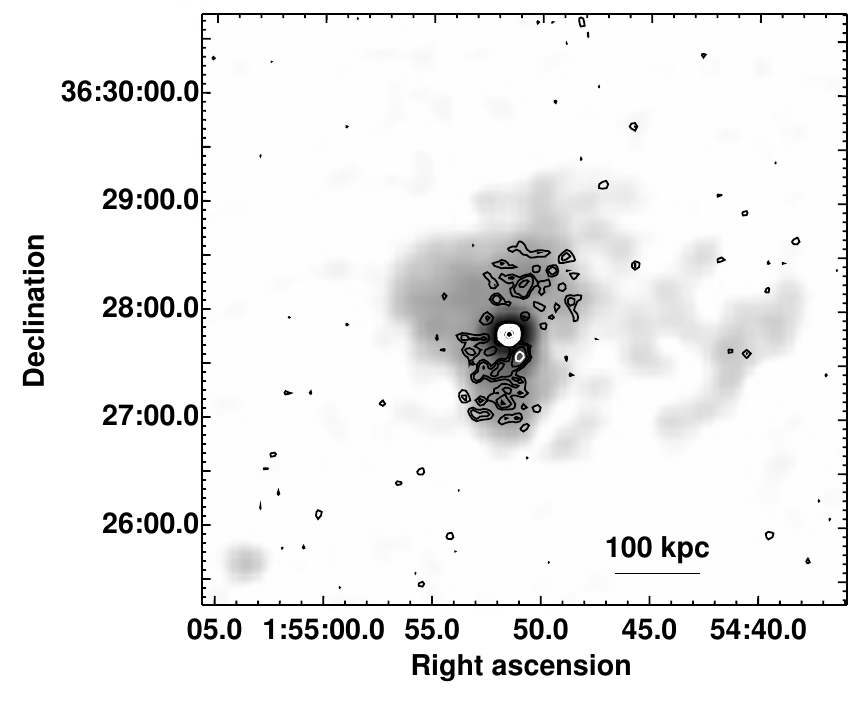}
\caption{VLA 4.7 GHz 
contours on the 1.4 GHz grey scale image of J\,0154+3627. The
resolution is 4.9$^{\prime \prime}\times$4.2$^{\prime \prime}$, p.a.
$-80$ and 15.8$^{\prime \prime} \times 13.4^{\prime \prime}$, p.a. $-74$.
Contours are spaced by a factor 2, starting from $\pm3\sigma$ (
1$\sigma=45$ $\mu$Jy b$^{-1}$ and 1$\sigma=50$ $\mu$Jy b$^{-1}$).
For this source the scale is 2.127 kpc/$^{\prime \prime}$.}
\label{fig:diffuse2}
\end{figure}

We searched the radio archives for available observations of this source 
at different frequencies. We found two Very Large Array (VLA)  
observations at 1.4 GHz and 4.9 GHz (both in the C--array configuration; 
project AC\,483), that we re--analysed. In Fig.~\ref{fig:diffuse2} 
we show the 4.9 GHz contours ($\sim$5$^{\prime \prime}$ resolution)
superposed to the grey scale image at 1.4 GHz ($\sim$16$^{\prime \prime}$ 
resolution). At 4.9 GHz the extended emission is mostly elongated along the 
North--South axis, consistent with the inner part of the 327 MHz image.
The observed morphology at 1.4 GHz is very similar to the images at 327 MHz
(Fig.~\ref{fig:diffuse}). An elongated region of very low brightness 
emission is detected on the western side of the source, where the low 
resolution image at 327 MHz reveals the existence of two radio tails. 

The total flux density at 327 MHz, measured on the low resolution image
in Fig.~\ref{fig:diffuse}, is 620.5$\pm31.0$ mJy. From the images in 
Fig.~\ref{fig:diffuse2}, we obtained a flux of 234.4$\pm9.4$ mJy at 1.4 GHz 
and 132.8$\pm5.3$ mJy at 4.9 GHz. 
The total spectral index is $\alpha=0.66^{+0.7}_{-0.6}$ between 327 MHz and 
1.4 GHz, and $\alpha=0.57^{+0.3}_{-0.4}$ between 327 MHz and 4.9 GHz. 
The radio information is summarised in Tab.~\ref{tab:wats_2}.

The compact component accounts for $\sim$3\% only of the total 
flux at 327 MHz (full resolution), while at higher frequency its 
contribution increases to $\sim$6\% at 1.4 GHz and $\sim$7\% at 4.9 GHz. 
Its spectral index is flat ($\alpha$=0.2) over the whole frequency 
range, indicating that this component hosts the radio core of 
J\,0154+3627. The spectral index of the diffuse emission is $\alpha$=0.6 
in the 327 MHz--4.9 GHz interval.

\begin{figure}
\centering
\includegraphics[angle=0,width=\hsize]{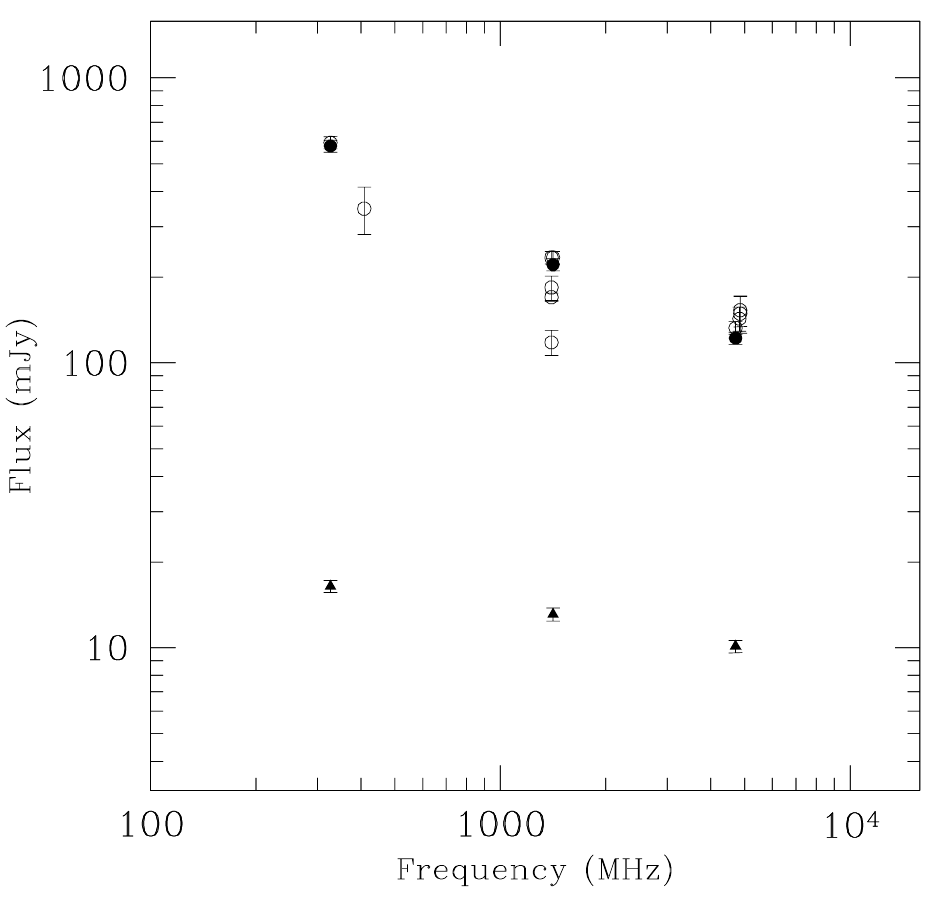}
\caption{Radio spectrum of the candidate tailed radio galaxy J0154+3627 
between 327 MHz and 4.9 GHz (empty circles). The spectrum of the core 
and extended emission are shown as filled triangles and filled circles, 
respectively.}
\label{fig:diffuse_sp}
\end{figure}

In Fig.~\ref{fig:diffuse_sp} we show the integrated radio spectrum of 
the source between 327 MHz and 4.9 GHz (empty circles). The spectrum was 
obtained by combining the fluxes in Tab.~\ref{tab:wats_2}, with the 
literature data provided by the NASA/IPAC Extragalactic Database (NED) 
for B2\,0151+36. The spread in the flux values at 1.4 and 4.9 GHz is most 
likely due to the different angular resolution of the observations. 
In the figure 
we also show the spectrum of the core (filled triangles) and of the diffuse 
component (filled circles). 

The morphological classification of this radio galaxy is not straightforward. 
The image shown in the left panel of Fig.~\ref{fig:diffuse} is reminiscent of 
3C\,317, classified as a core--halo radio galaxy (Zhao et al. 1993), a class 
of sources typically found at the centres of cool core clusters (e.g.
Baum \& O'Dea 1991; Sarazin et al. 1995;
Mazzotta \& Giacintucci 2008). However, the diffuse emission in core--halo
radio galaxies usually has a steep spectrum  ($\alpha \ge 1$), which is not 
observed here. Moreover, the western extension of the diffuse emission is 
suggestive of a tailed radio galaxy.

\section{Tailed radio galaxies as probes of the cluster environment}\label{sec:opt1}

We searched the literature to collect information about the environment 
of the new tailed radio galaxies presented in Sect.~\ref{sec:tailed}. 

We found that J\,0902+2044 and J\,0902+2045 (Fig.~\ref{fig:wats_1}) are located 
in the region of NSC\,J090232+204358 (hereinafter NSC\,J090+2043), 
a candidate galaxy cluster, on the basis of photometric redshifts, 
listed in the catalog of the Northern Sky Optical Cluster Survey (Gal et al. 
2003). Its photometric redshift is z$_{\rm phot}$=0.075$\pm$0.033, and its 
measured richness is N$_{\rm gal}$=39.2 (Tab.~\ref{tab:clusters}). 
\\
The WAT J\,1455+2209 (Fig.\ref{fig:wats_2}) is associated with the 
central galaxy of MaxBCG\,J223.97317+22.15620, 
classified as galaxy cluster by Koester et al. (2007) 
using the photometric information in the Sloan Digital Sky Survey.
For this cluster Koester et al. provide z$_{\rm phot}$=0.262$\pm$0.010
and N$_{\rm gal}$=18. Table \ref{tab:clusters} summarizes the general 
properties of the two candidate galaxy clusters.

We inspected the optical images from the SDSS to identify the host galaxies 
of the three radio sources. The optical counterparts of J\,0902+2044 and 
J\,0902+2045 are two elliptical galaxies with spectroscopic redshift 
z$_{\rm spec}$=0.083 and z$_{\rm spec}$=0.082, respectively\footnote{At these 
redshifts the linear scale is 1.56 kpc/$^{\prime \prime}$}. The agreement 
between these redshifts and the photometric redishift of the cluster, coupled 
with the tailed morphology of the two radio galaxies, confirms that 
NSC\,J090+2043 
is the cluster hosting them.
\\
The WAT J\,1455+2209 is associated with an elliptical at 
z$_{\rm spec}$=0.257\footnote{At this redshift the linear scale is 3.99 
kpc/$^{\prime \prime}$}. Also in this case the galaxy redshift is consistent, 
within the errors, with the photometric redshift of the 
cluster MaxBCG\,J223.97317+22.15620 (Tab.~\ref{tab:clusters}). 

The optical properties of the three tailed radio galaxies are summarised in 
Tab.~\ref{tab:wats_1}, where we provide the optical name and coordinates, 
spectroscopic redshift, and absolute red magnitude $\rm M_R$. All the optical 
information is taken from the latest release of the SDSS (Data Release 7). 

Using the redshift of the optical counterparts and the NVSS flux density 
in Tab.~\ref{tab:wats_1}, we calculated the radio power at 1.4 GHz. The 
WAT and the NAT in NSC\,J090+2043 have P$_{\rm 1.4 GHz}$= $2.2\times10^{24}$ 
W Hz$^{-1}$ and $1.3\times 10^{24}$ W Hz$^{-1}$, respectively. The radio power 
of the WAT J\,1455+2209 is P$_{\rm 1.4 GHz}$= $9.3\times10^{24}$ W Hz$^{-1}$ 
(Tab.~\ref{tab:wats_1}). 
The combination of these radio powers and the optical magnitude
of the associated galaxies, in the range M$_{\rm R}$ = --22.21 $\div$ --22.61
(see Table 2) places them in the transition region between FRI and FRII
radio galaxies in the  M$_{\rm R}$ -- logP$_{\rm 1.4GHz}$ diagram  
(Owen \& Ledlow,
1994). 
\\
The projected linear sizes of these three radio galaxies, in the range
100 -- 300 kpc (see Table 2), are consistent with what is typically found 
for tailed radio galaxies in clusters, whose sizes range over a wide 
interval, from tens to hundreds of kpc.

We inspected the X--ray images from the ROSAT All Sky Survey (RASS)
both for NCS~J\,0902+2043 and MaxBCG~J\,223.97317+22.15620. No X--ray
counterpart is detected in both cases. Such lack of detection may be
due to a number of reason. On one side, the number of optical galaxies 
provided in the literature (Tab.~\ref{tab:clusters}) suggests that they 
are poor clusters, and the lack of X--ray information might be due to 
the faint X--ray emission typical of these environments; on the other
side, at the redshift of MaxBCG~J\,223.97317+22.15620 the cluster 
X--ray emission might be below the sensitivity limit of the RASS. 
We note, for instance, that the rich and X--ray luminous 
(L$_{\rm X}=9.25\times10^{44}~$erg s$^{-1}$) merging cluster 
RXCJ\,2003.5--2323, located at z=0.3171, was barely detected on the RASS, 
and deep pointed Chandra observations were necessary for a radio/X--ray 
analysis (Giacintucci et al. 2009).

%
%
%
%

%
%
%
%

%
%
\begin{figure*}
\centering
\includegraphics[angle=0,width=17cm]{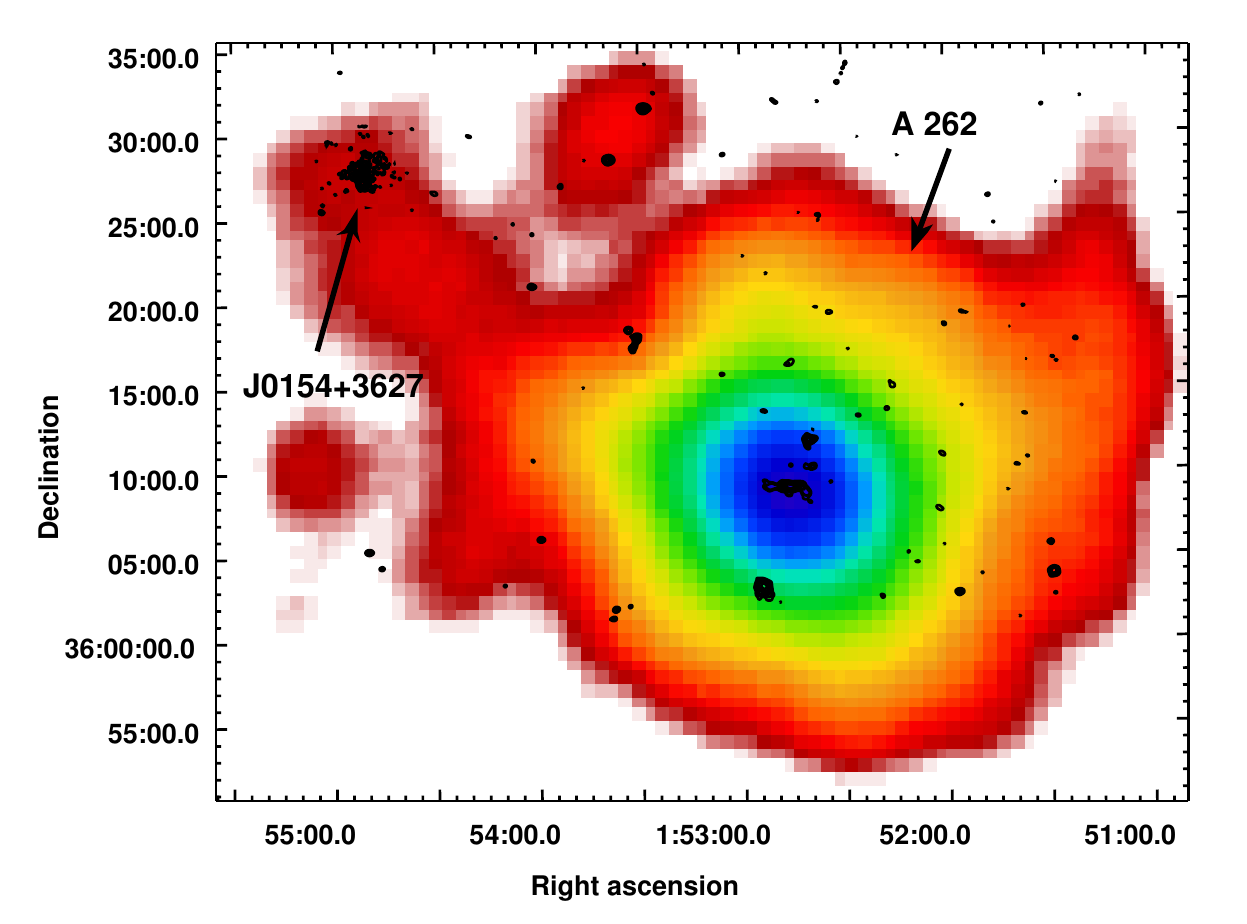}
\caption{GMRT 327 MHz contours, overlaid on the RASS image in the
0.5--2.1 keV band of the A\,262 region. The resolution of the radio 
image is 19.5$^{\prime \prime}
\times$15.2$^{\prime \prime}$, p.a. $-$86$^{\circ}$. The lowest contour
is at the 5$\sigma$ level=2.5 mJy b$^{-1}$, and each contours increases by a
factor of two. The RASS image has been smoothed with a gaussian
with $\sigma$=6$^{\prime}$.}
\label{fig:diffuse3}
\end{figure*}
%
%

\section{A galaxy cluster behind A\,262?}\label{sec:opt2}

As clear from Fig.~\ref{fig:diffuse} (left), the radio core
of the candidate tailed source J\,0154+3627 is coincident with a 
bright optical galaxy on the POSS--2 image. Its projected distance 
from the centre of the galaxy cluster A\,262 is $\sim$ 
32$^{\prime}$, which corresponds to $\sim$ 600 kpc at the
redshift of A\,262 (z=0.0163). 
However, its spectroscopic redshift, 
z$_{\rm spec}=0.118$\footnote{At this redshift the linear 
scale is 2.13 kpc/$^{\prime \prime}$} (Smith et al. 2004), rules 
out the membership to A\,262.

Using the flux density measured on the NVSS, and assuming a redshift of 
0.1177, we calculated the radio power of the source at 1.4 GHz. We 
obtained P$_{\rm 1.4 GHz}$= $8.5\times10^{24}$ W Hz$^{-1}$.
Also this radio galaxy is located in the FRI/FRII transition region 
in the M$_{\rm R}$ -- logP$_{\rm 1.4GHz}$ plot. Note that this galaxy is
the most luminous among those presented in this paper, even though it
is not as bright as the brightest central dominant galaxies in clusters
(M$_{\rm R} \sim -23$, Owen \& Ledlow 1994).
\\
It is difficult to provide an unambiguous morphological classification
for J\,0154+3627, since projection effects are likely to play a role.
The two short jets visible in Fig.~3, extending beyond the optical galaxy,
and the asymmetry of the diffuse emission (extended westwards) lead us
to think that we are dealing with a tailed source rather than 
with a core--halo radio galaxy.
The bright optical magnitude of the associated galaxy and the radio power 
would be consistent with a centrally located WAT, but the possibility
that we are dealing with a NAT radio galaxy cannot be
ruled out. In any case, we can safely conclude that J\,0154+3627 is a tailed
radio galaxy, most likely associated with a cluster at z=0.1177 located
behind A\,262.
\\
We checked on the X--ray archives in search for information in this band.
J\,0154+3627 is located outside the field of view of all the pointed 
observations of A\,262, and the only available image for this source
is provided by the RASS. Fig.~\ref{fig:diffuse3} shows the smoothed 
RASS image in the 0.5--2.1 keV band of A\,262 and the region of 
J\,0154+3627, with the GMRT contours at 327 MHz overlaid.
The image reveals a region of diffuse emission spatially coincident 
with the radio galaxy. Its projected distance 
from the centre of A\,262 is $\sim$600 kpc, and thus we cannot rule out 
the possibility that such emission is a clump of gas or an infalling group 
at the outskirts of the cluster, rather than emission coming from a
background cluster (or group). Pointed X--ray observations would be
useful to investigate the morphology of this emission and its possible
connection to the radio galaxy J\,0154+3627.

%
%

\begin{table*}[h!]\label{tab:clusters}
\label{tab:prop}
\caption[]{General properties of the candidate galaxy clusters.}
\begin{center}
\begin{tabular}{ccccc}
\hline\noalign{\smallskip}
Cluster name &  RA & DEC & z$_{\rm phot}$ & N$_{\rm gal}$ \\
   & (h,m,s)       & ($^{\circ}$, $^{\prime}$, $^{\prime \prime}$) & & \\
\hline\noalign{\smallskip}
NSC\,J090232+204358 & 09 02 32.7 & +20 43 57 & 0.075$\pm$0.033 & 39.2 \\
MaxBCG\, J223.97317+22.15620 & 14 55 53.5 & +22 09 22 & 0.262$\pm$0.010 & 18 \\
\noalign{\smallskip}
\hline
\end{tabular}
\end{center}
\end{table*}


\section{Summary and conclusions}\label{sec:summary}

In this paper we report on the serendipitous discovery of four radio 
galaxies, whose tailed radio morphology is suggestive of a cluster  
environment. Such discovery was possible thanks to the high sensitivity 
and wide field of view of the GMRT at 610 MHz (Venturi et al. 2007, 2008) 
and at 327 MHz. 

We searched for the optical counterparts of the new tailed 
radio galaxies, and found that J\,0902+2044 and J\,0902+2045, respectively a 
WAT and a NAT, are located in the candidate galaxy cluster NCS\,J0902+2043 
with photometric redshift 0.075. Our radio images thus confirm that 
NCS\,J0902+2043 is indeed a galaxy cluster, whose redshift is consistent 
with the radio power derived for the two radio galaxies. 

Similarly, we confirm that MaxBCG\,J223.97317+22.15620 (z$_{\rm phot}$=0.262) 
is a galaxy cluster, which hosts the central WAT J\,1455+2209.

The optical counterpart of the candidate tailed source J\,0154+3627 is 
a galaxy at redshift z=0.118. This rules out the possibility that the radio 
galaxy is associated with A\,262 (z=0.016). The observational properties of 
the radio source, i.e., morphology, spectrum, radio power and size, suggest 
that it is member of a galaxy cluster, which is so far undetected in the 
optical band. Inspection of the X--ray RASS image reveals a clump of 
diffuse emission coincident with the radio galaxy. However, the available
X--ray data do not allow us to tell whether such emission is related to the 
environment of J\,0154+3627 (z=0.118), or to the closer A\,262 (z=0.016).

Our findings confirm that distorted radio galaxies can be used as tracers 
of galaxy clusters. The low observing frequencies available with the 
GMRT (few hundred of MHz), coupled with its large field of view
(primary beam $\sim$ 1 to 2.5 degrees going from 610 MHz to 240 MHz),
very good sensitivity (from few tens to few hundreds of $\mu$Jy going from 
610 MHz to 240 MHz) and angular resolution (of the order of 5$^{\prime\prime}$
and 15$^{\prime\prime}$ respectively at 610 MHz and 240 MHz) 
are an ideal combination for the discovery of galaxy clusters
through tailed radio galaxies, not only in the Local Universe but also
at intermediate redshifts. We point out that the sensitivity of the REFLEX
survey (B{\"o}hringer et al. 2004) does not allow to detect any cluster 
with X--ray luminosity
L$_{\rm X} \le 5\times10^{44}$ erg s$^{-1}$ beyond z$\sim$ 0.1, therefore
the morphological properties of radio galaxies are the only observational
signature to confirm the existence of gravitationally bound systems above
such redshifts. Such result is relevant not only in itself, but 
also in the light of studies of the evolution of the radio galaxy population
in clusters of galaxies.
\\
\\
{\it Acknowledgements.}
We thank Prof. D. Dallacasa for insightful discussion and careful reading
of the manuscript. We thank the staff of the GMRT for their help during 
the observations. GMRT is run by the National Centre for Radio Astrophysics 
of the Tata Institute of Fundamental Research. We acknowledge financial 
contribution 
from the Italian Ministry of Foreign Affairs, from MIUR grants PRIN2004,
PRIN2005 and 2006, from PRIN--INAF2005 and from contract 
ASI--INAF I/023/05/01.

\end{document}